\documentclass[useAMS,usenatbib]{mn2e}
\usepackage{amsmath}
\usepackage{amssymb}
\usepackage{threeparttable}
\usepackage{cases}
\usepackage{color}
\usepackage{graphicx}
\usepackage[hyperfootnotes=false]{hyperref}
\usepackage[figure]{hypcap}

\hypersetup{colorlinks, citecolor=blue, pdfduplex=DuplexFlipLongEdge}
\hypersetup{pdftitle=Convective Accretion Discs, pdfauthor=Yucong Zhu, pdfsubject=Astrophysics}

\newcommand{\msun}{ M_{\odot}}

\newcommand{\grad}{\nabla}
\newcommand{\grada}{\nabla_{\rm a}}
\newcommand{\grade}{\nabla_{\rm e}}
\newcommand{\gradr}{\nabla_{\rm r}}

\title[Thermal Stability in Turbulently Mixed Accretion Discs]
  {Thermal Stability in Turbulent Accretion Discs}
\author[Y. Zhu \& R. Narayan.]
  {Yucong Zhu$^1$\thanks{E-mail:\newline \hbox{yzhu@cfa.harvard.edu (YZ);} \hbox{rnarayan@cfa.harvard.edu~(RN);}},
 Ramesh Narayan$^1$\footnotemark[1],
\\
  $^1$Harvard-Smithsonian Center for Astrophysics, 60 Garden Street, Cambridge, MA 02138, USA}
\date{\today}

\begin{document}

\maketitle

\label{firstpage}


\begin{abstract}
The standard thin accretion disc model predicts that discs around stellar mass black holes become radiation pressure dominated and thermally unstable once their luminosity exceeds $L \gtrsim 0.02 L_{\rm Edd}$.  Observationally, discs in the high/soft state of X-ray binaries show little variability in the range $0.01 L_{\rm Edd} < L < 0.5 L_{\rm Edd}$, implying that these discs in nature are in fact quite stable.  In an attempt to reconcile this conflict, we investigate one-zone disc models including turbulent and convective modes of vertical energy transport.  We find both mixing mechanisms to have a stabilizing effect, leading to an increase in the $L$ threshold up to which the disc is thermally stable.  In the case of stellar mass black hole systems, convection alone leads to only a minor increase in this threshold, up to $\sim5$ per cent of Eddington.  However turbulent mixing has a much greater effect -- the threshold rises up to $20$ per cent Eddington under reasonable assumptions.  In optimistic models with superefficient turbulent mixing, we even find solutions that are completely thermally stable for all accretion rates.  Similar results are obtained for supermassive black holes, except that all critical accretion rates are a factor $\sim$10 lower in Eddington ratio.

\end{abstract}

\section{Introduction}\label{sec:intro}

Accretion discs are ubiquitous in our universe; their physics governs the production of powerful jets from Active Galactic Nuclei, the formation of planets, and the growth of black holes.  This wealth of applicability has led to a rich field of study where many open problems still persist to this day \citep{frank2002}.  

One longstanding puzzle relates to the stability of discs.  In the standard $\alpha$-model for thin accretion discs \citep{shakura73,NT}, there is a critical accretion rate above which the disc transitions from being gas pressure dominated to radiation pressure dominated, and simultaneously switches from being thermally stable to unstable \citep{shakura76,piran78}.  The model predicts that the transition should occur at an accretion rate above a few tenths of a per cent of Eddington, depending on the mass of the central object.  Above this rate, we expect limit-cycle behaviour due to the onset of both thermal instability \citep{shakura76,honma92,szuszkiewicz98,janiuk02} and viscous instability \citep{lightman74}.  However, observationally there is little variability and no evidence of a limit-cycle for systems with substantially larger luminosities than the theoretical limit. Indeed, \citet{gierlinski04} show that discs around stellar mass black holes remain stable up to 50 per cent Eddington, which conflicts with the prediction of standard thin-disc theory by more than an order of magnitude.

Over the years, many ideas have been proposed to resolve this inconsistency.  \citet{piran78} noted that the prediction of thermal instability hinges crucially on taking the standard $\alpha$-viscosity prescription in disc models.  Other prescriptions for viscosity, such as having disc stress scale with gas pressure instead of total pressure, admit solutions that are thermally stable everywhere \citep{kato08}.  However, shearing box disc simulations with radiation \citep{hiroseblaes09} demonstrate that the standard $\alpha$-prescription, in which the stress scales linearly with the total pressure, is correct.  Thus, we do not have freedom to modify the viscosity prescription.

Other attempts to resolve the stability paradox in the context of the $\alpha$-viscosity paradigm include modeling discs with: strong irradiation at the disc surface\citep{tuchman90}; a superefficient corona that rapidly siphons heat from the disc and avoids instability by keeping the disc cool \citep{svensson94,rozanska99}; strong magnetic pressure support that dominates the vertical structure at low accretion rates\citep{zheng11}; time lags between the generation of disc stress and pressure response \citep{ciesielski12} as suggested by numerical simulations \citep{hirosekrolik09}; and finally convection \citep{milsom94,goldman95}.

In early works on convective discs, \citet{bisnovatyi77} and \citet{goldman95} found convection to be superefficient, dominating over radiation in the vertical transport of energy.  This superefficient convective flux strongly suppresses the radiative channel, resulting in cooler discs that are completely thermally stable \citep{goldman95}.  However, other groups who performed more detailed calculations in which they evaluated the complete vertical structure \citep{shakura78,cannizzo92,milsom94,heinzeller09} found a more modest effect where convection carries no more than $\sim 1/2$ of the vertical flux of energy, and has only a weak effect on the disc's stability \citep{cannizzo92, milsom94, sadowski11}.  

These later models still find that convection does provide a stabilizing force that pushes the instability threshold towards higher accretion rates, but the effect is modest and falls far short of what is needed to explain observations.  In this connection, we note that turbulent convection alone is too small by a factor of 10-100 \citep{ruden88,ryu92,goldman95} to produce the viscous stress present in accretion discs \citep{pringle72,king07}.

Disc viscosity is believed to arise from the Magneto-Rotational Instability (MRI; \citealt{balbus91,balbus98}).  One consequence of MRI turbulence is that it will itself induce a vertical transfer of energy, in a somewhat analogous fashion as convection.  What effect will this have on the thermal stability of the disc?  To date, no analytic model has been developed that takes into account both convection and turbulence.  In this work we attempt to develop the simplest version of such a model.  Based on the results of \citet{agol01}, we expect turbulent mixing to have a significant impact on the disc's vertical structure and hence also on disc stability.  We find that turbulence indeed acts in a similar way to convection, i.e. turbulent discs are substantially more resilient to the onset of thermal instability.  For reasonable choices of model parameters, we show that it is possible to push the instability threshold up close to the observational limits.  

The organization of the paper is as follows.  In \S\ref{sec:setup}, we present the governing equations and details of our turbulent disc model.  We then proceed to compute three types of discs using our model: 1) a classic purely radiative disc, 2) a convective disc, and 3) a convective plus turbulent disc.  In \S\ref{sec:results} we compare the stability properties and radial structure for these three types of disc.  Section \ref{sec:discussion} focuses on the impact of various assumptions made in our disc model and also briefly compares our work to the results of numerical simulations. Finally, we conclude in \S\ref{sec:conclusion} with a summary of the important points.



\section{Physical model}\label{sec:setup}

We wish to build the simplest possible disc model that includes the physics of convection and turbulent vertical mixing.  To this end, we separate the equations describing the disc's vertical and radial structure, and model the vertical structure as a single homogeneous slab (see Fig. \ref{fig:schematic}).  In essence, our approach is similar to the one-zone relativistic disc model of \citet{NT}, but including the effects of convective and turbulent mixing.

The primary effect of convection and turbulence is to add a new channel for vertical energy transport.  Since the computation of convective flux requires information about the vertical gradient of temperature, any convective model must have at least two vertically separated probe points.  For convenience, we choose these two sampling points to be the disc midplane and the density scale height.\\

The strategy we employ is to first obtain as a function of radius quantities that are independent of the vertical disc structure.  We then use these quantities as boundary conditions to uniquely specify the disc vertical structure, and hence solve for the complete disc model.

\subsection{Radial structure}

\subsubsection{Luminosity:}  For simplicity, we assume that the advected energy is negligible.  This allows us to calculate the net radiative flux leaving the disc at each radius, given only the black hole mass $M$, dimensionless spin $a_* = J/(GM^2/c)$, and mass accretion rate $\dot{M}$.  We adopt the prescription of \citet{pagethorne74} for determining the disc's luminosity profile.  In the subsequent vertical structure equations, we make use of the total emergent disc flux

\begin{equation}
F_{\rm tot} = \sigma T_{\rm eff}^4.
\end{equation}

\subsubsection{Vertical gravity:}  The tidal vertical gravity in the disc is a function purely of the Kerr metric and the motion of the disc fluid.  For circular orbits, the tidal vertical gravity $g_z$ is given by \citep{riffert95}:
\begin{equation}\label{eq:Qdef}
g_z \equiv Q\cdot z = (GM/r^3)^{1/2} R_{z}(r) \cdot z,
\end{equation}
where $z$ is height above the midplane, and $R_{z}(r)$ is a dimensionless relativistic factor given by:
\begin{equation}\label{eq:RiffertHerold}
R_{z}(r) = \frac{1-4a_*/r_*^{3/2}+3a_*^2/r_*^2}{1-3/r_*+2a_*/r_*^{3/2}}.
\end{equation}
We have defined the dimensionless radius $r_* = r/(GM/c^2)$.  Note that the quantity $Q=g_z/z$ is independent of $z$.\\

\subsubsection{Rotation rate:}  For circular motion about a Kerr black hole, the orbital velocity is given by (c.f. \citealt{NT}):
\begin{equation}
\Omega_k = \left(\frac{GM}{r^3}\right)^{1/2} \left(\frac{1}{1+a_*/r_*^{3/2}}\right).
\end{equation}

\subsection{Vertical structure}

\subsubsection{Unknown variables} \label{sec:unknowns}

At every fixed radius of the accretion disc, we wish to solve for the following 8 unknowns that describe the vertical structure: 

\begin{itemize}
\item $T_{\rm mid}$ -- Midplane temperature
\item $P_{\rm mid}$ -- Midplane pressure (gas+radiation)
\item $T_{\rm scale}$ -- Temperature at one scale height
\item $P_{\rm scale}$ -- Pressure at one scale height (gas+radiation)
\item $\Sigma$ -- Vertical column density
\item $H$ -- Vertical pressure scale height
\item $\grad = d\ln{T}/d\ln{P}$ of the ambient medium
\item $\grade = d\ln{T}/d\ln{P}$ of a convective element
\end{itemize}

\subsubsection{Equations:} \label{sec:verticalEquations}
We employ the following 8 equations to solve for the 8 unknowns:

\begin{enumerate}[(I)]
\item Vertical pressure balance
This gives
\begin{equation}
\frac{dP_{\text{tot}}}{dz} = \rho g_z , \nonumber
\end{equation}
which has the vertically integrated form:
\begin{equation}\label{eq:pressurebal}
\frac{P_{\text{tot}}}{H} = \rho Q H = \frac{\Sigma Q}{2},
\end{equation}
where $H$ is the vertical pressure scale height, $Q=g_z/z$ is defined by Eq.\eqref{eq:Qdef}, and we have written the density as $\rho = \Sigma/2H$ in the spirit of a one-zone model.

\item Viscous heating:

Through the energy equation for viscous heating, it is possible to link the disc flux $F_{\rm tot}$ with the vertically integrated stress $W$ (See 5.6.7-12 of \citealt{NT}).  The resulting expression is
\begin{subequations}\label{eq:heating}
\begin{equation}
F_{\text{tot}}=\frac{3}{4} \Omega_k R_F(r) W,
\end{equation}
with the dimensionless relativistic factor $R_F(r)$ defined as
\begin{equation}
R_F(r) = \frac{1-2/r_*+a_*^2/r_*^2}{1-3/r_*+2a_*/r_*^{3/2}}.
\end{equation}
Taking the $\alpha$-prescription for the stress (where $t_{\hat{\phi}\hat{r}}=\alpha P_{\text{tot}}$), we have:
\begin{equation}
W \equiv \int_{-H}^{H} t_{\hat{\phi}\hat{r}} dz = 2 H \alpha P_{\rm mid}.
\end{equation}
\end{subequations}
\\

\item Midplane equation of state:

We ignore magnetic pressure in this analysis, and adopt an ideal gas plus radiation equation of state,
\begin{align}\label{eq:midplaneEOS}
P_{\text{mid}} & =\frac{\rho k_B T_{\rm mid}}{\mu} + \frac{aT_{\rm mid}^4}{3} \nonumber \\
& = \frac{\Sigma k_B T_{\rm mid}}{2H\mu} + \frac{aT_{\rm mid}^4}{3},
\end{align}
where $k_B$ is the Boltzmann constant, and $\mu$ is the mean molecular weight of the fluid (taken to be 0.615$m_{\rm H}$, which corresponds to ionized gas comprised of 70 per cent Hydrogen and 30 per cent Helium by mass).
\\
\item Equation of state at a density scale height:

We define the density scale height to be where the density falls to $e^{-1}$ of its midplane value.  Thus the equation here is:
\begin{equation}\label{eq:scaleEOS}
P_{\text{scale}}=\frac{\Sigma k_B T_{\rm scale}}{2 e H\mu} + \frac{aT_{\rm scale}^4}{3}.
\end{equation}
\\

\item Radiative diffusion:

By integrating the second moment of the radiative transfer equation $dP/d\tau=F_{\rm rad}$ and using the condition of constant radiative flux, we arrive at the following expression for the vertical temperature profile (for a grey-atmosphere):
\begin{subequations}
\begin{equation}
T(\tau) = \left[\frac{3}{4}\left(\frac{F_{\rm rad}}{\sigma}\right)\left(\frac{2}{3} + \tau\right)\right]^{1/4}, \label{eq:radiativediffusion}
\end{equation}
where $\tau$ is the optical depth measured from the surface, $\sigma$ is the Stefan-Boltzmann constant, and $F_{\rm rad}$ is the radiative flux as evaluated from the radiative diffusion equation:
\begin{align}
F_{\rm rad} & = \frac{4}{3}\frac{\sigma}{\rho \kappa} \frac{dT^4}{dz} \label{eq:radfluxNT} \\
 & \approx \left(\frac{4ac}{3}\frac{QHT_{\rm mid}^4}{\kappa P_{\rm mid}}\right) \cdot \grad . \label{eq:radflux}
\end{align}
\end{subequations}
Equation \eqref{eq:radiativediffusion} gives the temperature at the scale height $T_{\rm scale}$ by plugging in $\tau_{\rm scale} = \kappa \Sigma_{\rm scale}$ for the optical depth, where $\Sigma_{\rm scale}$ refers to the vertical column density from the scale height to the disc surface, and $\kappa$ is the opacity, which for simplicity we take to be the electron scattering opacity $\kappa_{es}$. The latter is justified since we are dealing with fairly hot discs.  Now, we need a way to estimate $\Sigma_{\rm scale}$, defined as the location where the density falls from its mid-plane value by an e-fold (cf. Eq. \ref{eq:scaleEOS}).  In our highly simplified one-zone model, we make the approximation that $\Sigma_{\rm scale} = 0.1 \Sigma$, which is roughly the value measured in more detailed multi-zone treatments of the vertical structure (see \S\ref{sec:sigmaScale} for more discussion).\\

\item Thermodynamic gradient $\grad$:

We calculate $\grad$ directly from the values of $P_{\rm mid}, T_{\rm mid}, P_{\rm scale}, T_{\rm scale}$ via:

\begin{equation}\label{eq:grad}
\grad = \frac{d\ln{T}}{d\ln{P}} \approx \left(\frac{\ln T_{\rm mid} - \ln T_{\rm scale}}{\ln P_{\rm mid} - \ln P_{\rm scale}}\right).
\end{equation}
\item Energy transport:

Radiation, convection, and turbulence all contribute to the vertical transport of energy.  Thus, the expression for the total energy flux $F_{\rm tot}$ is:
\begin{subequations}\label{eq:convectivediffusion}
\begin{equation}\label{eq:fluxconservation}
F_{\rm tot} = F_{\rm rad} + F_{\rm conv} + F_{\rm turb},
\end{equation}
where $F_{\rm rad}$ is the radiative flux given by Eq. \eqref{eq:radflux}, and $F_{\rm conv}$ is the convective flux given by \citep{mihalas78}:
\begin{equation}\label{eq:fconv}
F_{\rm conv} = \rho c_{\rm p} \bar{v}_{\rm convect} \Delta T.
\end{equation}
Here $c_{\rm p}$ is the specific heat capacity, $\bar{v}$ is the average vertical speed of convective blobs, and $\Delta T$ is the typical temperature differential between a fluid parcel and the ambient medium.  We evaluate these quantities using standard mixing-length theory, with mixing length $\Lambda$ taken to be some multiple of the pressure scale height $H$.  As a default we set $\Lambda/H = 1$, though we also consider other values.  The convective mixing-length velocity is then \citep{mihalas78}:
\begin{equation}\label{eq:vconv}
 \bar{v}_{\rm convect} = \left|\frac{Q\Lambda^2}{8}(\grad - \grade)\right|^{1/2}.
\end{equation}
Henceforth, we use subscript-e to denote quantities that are measured within the convective blob;  $\grade$ therefore represents the effective $d\ln T/d\ln P$ experienced by a convective/turbulent element as it moves vertically before dissolving back into the surrounding medium.  For the temperature differential, we take:
\begin{align}
\Delta T & = \Lambda \left(\frac{dT}{dz} - \left. \frac{dT}{dz}\right|_{\rm e}\right) \nonumber \\
& = \frac{\Lambda T}{P} \left(\frac{dp}{dz}\right) \left(\frac{d\ln{T}}{d\ln{P}} - \left. \frac{d\ln{T}}{d\ln{P}}\right|_{\rm e}\right) \nonumber \\
& = \frac{\Lambda T_{\rm mid}}{H} \left(\grad - \grade\right).
\end{align}
Here, the specific heat capacity $c_{\rm p}$ is given by standard formulae corresponding to a monatomic gas/radiation mixture \citep{chandrasekhar67} using midplane quantities to set the gas/radiation pressure ratio.

For turbulent mixing, we follow the same prescription as our convective mixing length theory (Eq. \ref{eq:fconv}).  We assume that the turbulent mixing flux is given by:
\begin{align}\label{eq:fluxturb}
F_{\rm turb}  & = \rho c_{\rm p} \bar{v}_{\rm turb} \Delta T \\ \nonumber
& =\rho c_{\rm p} ( \bar{v}_{\rm turb} \Lambda_{\rm turb} ) \frac{T_{\rm mid}}{H}(\grad -\grade),
\end{align}
where we now invoke a turbulent velocity $\bar{v}_{\rm turb}$ and turbulent scale height $\Lambda_{\rm turb}$.   To estimate the size of the product $(\bar{v}_{\rm turb} \Lambda_{\rm turb})$, recall that turbulent viscosity has the form
\begin{equation}\label{eq:turbvisc}
\nu_{\rm turb} \sim \bar{v}_{\rm turb} \Lambda_{\rm turb}.
\end{equation}
Consistency with the $\alpha$-prescription requires that this viscosity equal
\begin{equation}\label{eq:alphavisc}
\nu_{\rm turb} = \alpha c_{s} H,
\end{equation}
where $\alpha$ is the viscosity coefficient introduced in Eqs. \eqref{eq:heating}, $c_s = \sqrt{2HP_{\rm mid}/\Sigma}$ is the fluid sound speed, and $H$ is the disc scale height.  Comparing Eqs. \ref{eq:turbvisc} and \ref{eq:alphavisc} and further assuming that $\Lambda_{\rm turb} = H$, we arrive at $\bar{v}_{\rm turb}=\alpha c_{s}$.  For the purpose of exploration, we allow the following more general scaling:
\begin{equation}
 \bar{v}_{\rm turb} = \zeta c_s,\label{eq:turbVelocity}
\end{equation}
where $\zeta$ is a dimensionless number $< 1$.  For most of our models, we set $\zeta=\alpha$ as our fiducial value.  A detailed investigation of how $\zeta$ affects the disc solution is presented in \S\ref{sec:zeta}.\\
\end{subequations}

\item Effectiveness of convection and turbulence vs. radiative diffusion:

To complete our mixing-length theory, we require another set of equations relating $\grad$ and $\grade$.  This is done by comparing the efficiency of convective and turbulent transport with radiative diffusion:
\begin{subequations}\label{eq:gamma}
\begin{equation}
\gamma \equiv \frac{\rm convective/turbulent \;energy\;lost\;at\;dissolution\;of\;blob}{\rm radiative\;energy\;lost\;during\; lifetime\;of\;blob}.
\end{equation}
The total energy loss from the convective and turbulent element is proportional to ($\grad - \grada$), where $\grada$ represents the adiabatic gradient of the surrounding fluid, i.e. $d\ln T/d\ln P$ for an ideal gas/radiation mixture.  It is given by standard formulae \citep{chandrasekhar67} and only depends on the gas/radiation pressure ratio; $\grada = 1/3$ in the gas limit and $\grada = 1/4$ in the radiation limit.  The total energy loss from the blob can be split into two components: 1) the energy released by dissolution at the end of the blob's life (proportional to $\grad - \grade$), and 2) the radiative energy loss from radiative diffusion out of the blob before it dissolves (proportional to $\grade - \grada$).   From \citet{mihalas78} p.189-190, the ratio of these two components $\gamma$ can be written as:
\begin{align}\label{eq:gammafull}
\gamma & = \frac{\grad - \grade}{\grade - \grada} \\ \nonumber
& = \left(\frac{\Sigma c_p \left(\bar{v}_{\rm convect} + \bar{v}_{\rm turb}\right)}{16 H \sigma T_{\rm mid}^3}\right) \left(\frac{1 + \tau_m^2/2}{\tau_m}\right),
\end{align}
where $\tau_m$ represents the optical depth of the convective cell, which we take to be:
\begin{equation}
\tau_m = \kappa \Sigma/2
\end{equation}
\end{subequations}
\end{enumerate}

\subsubsection{Solving the system of equations}

Eqs. \eqref{eq:pressurebal}-\eqref{eq:gamma} form a system of 8 equations, where the only unspecified values are the 8 unknowns listed in \S\ref{sec:unknowns}.  These 8 equations can be solved at each radius in the disc to yield a complete model.  The set of model parameters needed are: 1) the central black hole mass $M$, 2) the black hole spin $a_*$, 3) the accretion rate $\dot{M}$, 4) the viscosity parameter $\alpha$, and 5) the two mixing parameters $\Lambda$, $\zeta$.  Although the system of equations is highly non-linear, Appendix \ref{app:solvingEqs} describes a simple procedure to numerically solve for all unknowns.  We discuss numerical results in the following section.

\section{Disc solutions}\label{sec:results}

\begin{figure}
\begin{center}  
\includegraphics[width=0.5\textwidth]{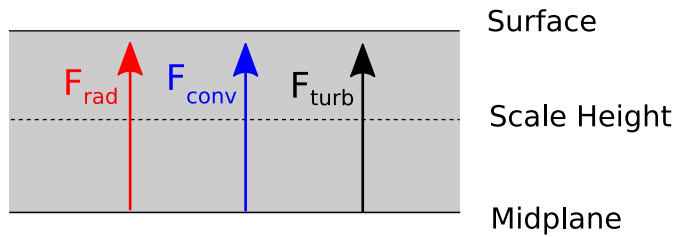}
\caption{A schematic of the vertical structure in our one-zone model.  We assume a homogeneous disc where energy is transported vertically along three channels: radiative diffusion ($F_{\rm rad}$), convection ($F_{\rm conv}$), and additional turbulent mixing ($F_{\rm turb}$).  We assume that all heating occurs  at the disc midplane so that $F_{\rm rad}$, $F_{\rm conv}$, and $F_{\rm turb}$ are constant everywhere.  We compute vertical gradients by comparing values at the midplane and at a density scale height.  In this work, we compare three distinct classes of disc models: a) purely radiative discs (with only $F_{\rm rad}$), b) convective discs (with $F_{\rm rad}$ and $F_{\rm conv}$), and c) turbulently mixed convective discs (including all 3 components $F_{\rm rad}$, $F_{\rm conv}$, and $F_{\rm turb}$). \label{fig:schematic}}  
\end{center}  
\end{figure}

Using the methods outlined in \S\ref{sec:setup} and Appendix \ref{app:solvingEqs}, we have calculated a wide array of disc models to understand how convection and turbulent mixing affect the overall disc structure.  We compare three distinct classes of disc models (see Fig. \ref{fig:schematic} for a schematic): a) Purely radiative discs with no convective/turbulent mixing (these correspond to the classic solutions of \citealt{NT} and are identified as ``No mixing'' in the plots);  b) Convective discs with vertical energy transport via both radiation and convection (these models are labelled as ``Convect'');  c) discs with vertical energy transport via radiation, convection, and turbulent mixing (labelled as ``Conv+Turb'').  A direct comparison of the stability properties for the 3 classes of disc models is shown in Fig. \ref{fig:scurveSpin} and discussed below in \S\ref{sec:turbdisc}.  For all cases, we have calculated models spanning accretion rates over a wide range, $0.001 < \dot{M}/\dot{M_{\rm Edd}} < 1$, where we define the Eddington accretion rate to be:
\begin{equation}
\dot{M}_{\rm Edd} = \frac{4\pi G M}{\kappa_{es} c \epsilon}.
\end{equation}
Here $\epsilon = 1 - E_{\rm ISCO}$ is the accretion efficiency of the disc, a measure of the total gravitational energy released by matter along its journey to the ISCO.  The efficiency $\epsilon$ ranges from $\sim 6-40$ per cent depending on black hole spin $a_*$.  Unless specifically noted otherwise, models were calculated with the following fiducial parameters : black hole mass $M=10\msun$, black hole spin $a_*=0$, disc radius $r_*=12$ (twice the ISCO radius for $a_*=0$), viscosity coefficient $\alpha=0.1$.  For models with convection (models ``Convect'' and ``Conv+Turb''), the mixing length was set equal to the scale height $\Lambda = H$.  Finally, in the case of turbulent mixing (model ``Conv+Turb''), the mixing velocity was set to $\bar{v}_{\rm turb} = \alpha c_s$, i.e. $\zeta=\alpha$.

\subsection{Classic unmixed disc}

To set the stage, we first solve our disc equations without invoking either convective or turbulent flux transport.  This corresponds to the classic disc solution of \citealt{NT}, where at each radius, we only need to solve for 4 unknowns: $T_{\rm mid}, P_{\rm mid}, \Sigma$, and $H$.  We obtain the disc model by solving a reduced set of 4 equations: Eq. \eqref{eq:pressurebal}, Eq. \eqref{eq:heating}, Eq. \eqref{eq:midplaneEOS}, and Eq. \eqref{eq:radfluxNT}.  In Eq. \eqref{eq:radfluxNT}, we substitute $\rho=\Sigma/2H$ and write the differential $dT^4/dz$ as
\begin{equation}
\frac{dT^4}{dz} \approx \frac{T_{\rm mid}^4}{H}.
\end{equation}
\begin{figure}
\begin{center}  
\includegraphics[width=0.5\textwidth]{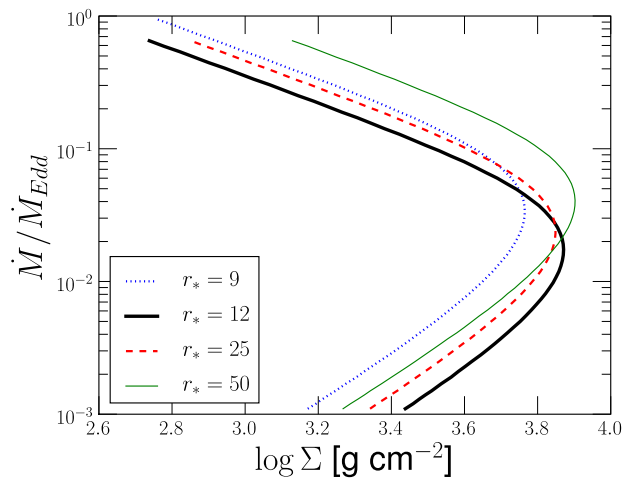}
\caption{Disc solutions in the $\dot{M}-\Sigma$ plane (``S-curve'') for a purely radiative disc with $\alpha=0.1$ around a non-spinning black hole.  The different tracks represent different disc radii measured in units of $r_* = r/(GM/c^2)$.  The slope $d\dot{M}/d\Sigma$ is an indicator for stability.  Solutions with $d\dot{M}/d\Sigma > 0$ are thermally and viscously stable whereas $d\dot{M}/d\Sigma < 0$ represents instability.  Thus, the characteristic bend in the solutions signifies the transition point from stable to unstable discs.  Among the four tracks shown, the lowest $\dot{M}$ for the bend occurs at around 2 per cent Eddington in the $r_*=12$ track.
\label{fig:scurveRadius}}
\end{center}  
\end{figure}  
The method used to solve this set of 4 equations is outlined in Appendix B of \citet{zhu12}.  In general, we find that for all choices of disc parameters $M, a_*, r_*, \alpha$, at any given radius the solutions fold back in the $(\dot{M},\Sigma)$ plane (Fig. \ref{fig:scurveRadius} shows examples).  It is well known that the slope of the solution track is an indicator for the disc's viscous and thermal stability \citep[e.g.][]{bath82, kato08}.  Solutions with $d\dot{M}/d\Sigma > 0$ are stable, while those with $d\dot{M}/d\Sigma < 0$ are unstable.  This result is independent of the details of the heating/cooling prescription.  Thus, the characteristic bend that we see in the $\dot{M}-\Sigma$ plane signifies that, at all radii, the accretion solution transitions from being stable at low $\dot{M}$ to unstable above some critical accretion rate.  Disc models that include the physics of advection \citep{abramowicz88, sadowski11} exhibit a second turnover at high $\dot{M}/\dot{M}_{\rm Edd}$ due to rapid advective cooling.  Above this $\dot{M}$, the disc is stable.  The resulting track in the $\dot{M}-\Sigma$ plane has an ``S'' shape.  Although our solutions do not capture this upper advection stabilized branch, for consistency with previous analyses on disc stability, we will still refer to our disc solution track in the $\dot{M}-\Sigma$ plane as an ``S-curve''.

The focus of the present paper is the lower two segments of the S-curve, and the transition from thermal stability at lower $\dot{M}$ to instability at higher $\dot{M}$.  The thermal instability threshold is located approximately where the disc transitions from being gas-pressure to radiation-pressure dominated.  When the disc is gas pressure dominated $P \sim P_{\rm gas} \sim T$, and for a fixed $\Sigma$, the cooling rate $Q_{\rm cool} \sim \sigma T_{\rm eff}^4 \sim T^4$ is a steeper function of temperature than the heating rate $Q_{\rm heat} \sim H t_{r\phi} \sim H \alpha P \sim P^2 \sim T^2$.  This implies thermal stability since any positive temperature perturbation is quickly eliminated as the system responds with net cooling.  However, the opposite occurs in the radiation dominated limit. Here, $P\sim P_{\rm rad} \sim T^4$, and while the cooling still scales as $Q_{\rm cool} \sim T^4$, the heating is steeper where $Q_{\rm heat} \sim P^2 \sim T^8$.   This is unstable since a positive temperature perturbation is self-reinforcing.  

It is also easy to understand the variation with radius of the critical $\dot{M}$ for stability.  The transition from being gas to radiation pressure dominated as one increases $\dot{M}$ first occurs in the most luminous parts of the disc.  This is indeed borne out in figure \ref{fig:scurveRadius}, where we see the lowest critical $\dot{M}$ occurs in the $r_*=12$ track, which lies closest to where the bulk of the disc energy is released ($dL/d\ln r$ is maximum at $r_*\sim 15$).

\subsection{Convective solutions}
\begin{figure}
\begin{center}  
\includegraphics[width=0.5\textwidth]{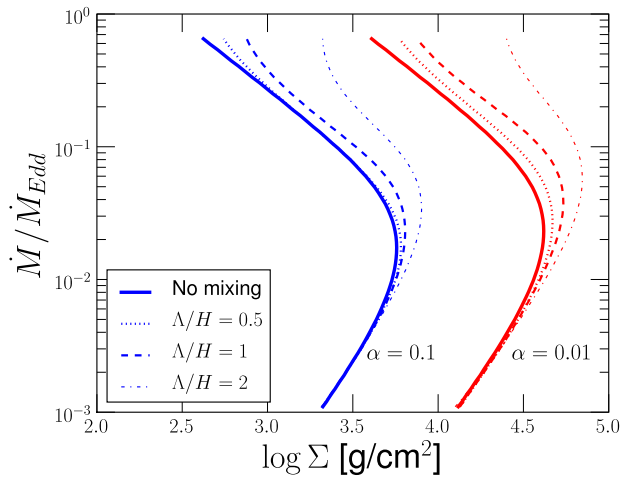}
\caption{Solution tracks in the $\dot{M}-\Sigma$ plane (``S-curve'') at $r_*=12$ for two choices of $\alpha=0.1$ (left), $\alpha=0.01$ (right), and three choices of the convective length scale:  $\Lambda / H =$ 0.5 (dotted), 1 (dashed), 2 (dash-dotted).  The classic nonconvective disc solution is shown by thick solid lines.  The net effect of increasing the mixing length is to increase the strength of convection, which tends to push the solutions out toward higher column densities.  Convective cooling also increases the critical turnoff accretion rate.  However, even with an optimistic choice of $\Lambda / H=2$, the critical turnoff point is no more than $\sim 5$ per cent Eddington for any of the convective solutions.  \label{fig:scurveAlpha}}  
\end{center}  
\end{figure}

To obtain the class of convective disc solutions, we solve the full system of equations outlined in \S\ref{sec:verticalEquations}, but without including a turbulent flux component (i.e. we set $\bar{v}_{\rm turb}=0$ in Eq.\ref{eq:fluxturb}).  As Fig. \ref{fig:scurveAlpha} shows, convective disc models are qualitatively very similar to the standard nonconvective discs.  Convection provides a modest stabilizing effect on the disc.  Therefore the transition value of $\dot{M}$ is a factor $\lesssim 2$ higher than in the corresponding purely radiative model (Fig. \ref{fig:scurveAlpha}).  Convection also causes the disc solutions to move towards higher column densities.  To understand this, note from Eqs. \eqref{eq:pressurebal} and \eqref{eq:heating}, that for a fixed $F_{\rm tot}$, we have $\Sigma\propto H^{-2}$ (cf. Appendix Eq. \ref{eq:Hscaling}).  Since convective discs are cooler than their nonconvective counterparts (e.g. only a fraction of the flux is carried by radiation, necessitating a smaller temperature gradient), they are thinner (smaller $H$) and hence have larger column densities. 
Another feature that we see in our convective disc solutions is that the onset of strong convection occurs strictly in the radiation dominated regime (in Fig. \ref{fig:scurveAlpha}, convection primarily modifies the upper, radiatively dominant branch).  This fact can be understood by the following argument: as the solution becomes more radiation dominated, the adiabatic gradient gets pushed to lower values (i.e. starting from $\grada \sim 1/3$ in the gas limit, the gradient falls to $\grada \sim 1/4$ in the radiation limit).  However, a disc becomes radiation dominated only at high accretion rates, and high accretion rates necessitate larger radiative gradients to push out the increased flux (cf. Eq. \ref{eq:radflux}, where higher $\dot{M}$ and hence higher $F_{\rm rad}$ implies larger $\grad$).  This divergence between adiabatic and radiative gradients (specifically, the push towards $\grad \gg \grada$) at high accretion rates is the engine that drives convection.  

Even given optimistic assumptions about convection (e.g. solutions with $\Lambda/H=2$ in Fig. \ref{fig:scurveAlpha}), we find that the instability threshold for convective discs stays well below 10 per cent Eddington.  To within a factor of 2, the instability threshold is similar to that found in nonconvective discs (compare solid with dashed lines in Fig. \ref{fig:scurveAlpha}).  Thus, we conclude that the action of convection alone results in only a modest stabilizing effect.

\subsection{Convective and turbulent disc solutions}\label{sec:turbdisc}

We now ask if MRI-induced turbulent mixing can provide a larger boost to the instability threshold.  Our treatment of the turbulent velocity is given by Eq.\eqref{eq:turbVelocity}.  For simplicity, we take as an ansatz that the proportionality constant $\zeta$ between the fluid sound speed and turbulent eddy velocity is $\zeta=\alpha$.  We  explore the impact of varying $\zeta$ in \S\ref{sec:zeta}.

\begin{figure}
\begin{center}  
\includegraphics[width=0.5\textwidth]{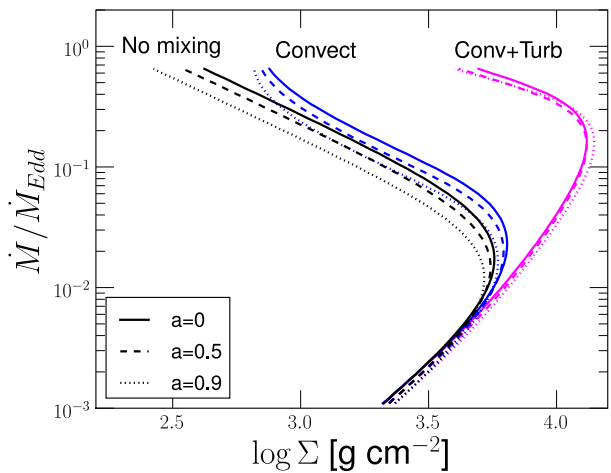}
\caption{S-curves for 3 choices of BH spin, calculated in each case at twice the ISCO radius: $r_*=(12,9,5)$ for spins $a_*=(0,0.5,0.9)$ respectively.  The three sets of curves correspond to the three classes of disc solutions: classic disc on the left (No mixing), convective disc in the centre (Convect), and convective + turbulent disc on the right (Conv+Turb).  Varying black hole spin does not have a significant impact on the location of the turnoff $\dot{M}$.  Note that the turnoff $\dot{M}$ in the turbulent models greatly exceed those in the other two disc models.  This shows that the stabilizing effect of turbulent mixing is much stronger than that of convective mixing alone.  The models assume $\Lambda/H=1$, $\alpha=0.1$, and $\zeta=\alpha$.\label{fig:scurveSpin}}  
\end{center}  
\end{figure}  

\begin{figure}
\begin{center}  
\includegraphics[width=0.5\textwidth]{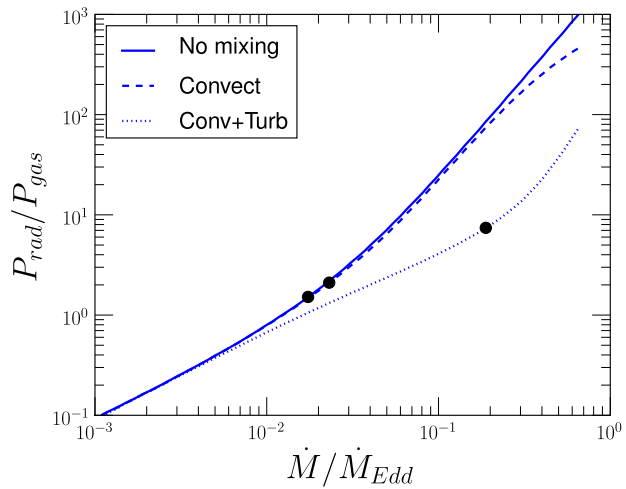}
\caption{Comparison of radiation-to-gas pressure ratios for $a_*=0$, $r_*=12$, and the three classes of disc solutions shown in Fig. \ref{fig:scurveSpin}.  The large black dots denote the critical accretion rate for each solution.  The pressure ratio for this critical point is not the same in the three classes of discs: $P_{\rm rad}/P_{\rm gas}=(1.5,2.1,7.3)$ for the radiative, convective, and turbulent disc models respectively.  Note also that we require higher $\dot{M}$ to reach the same $P_{\rm rad}/P_{\rm gas}$ in the convective solutions as in the ``no mixing'' case.\label{fig:P}}
\end{center}  
\end{figure}

Fig. \ref{fig:scurveSpin} shows our results.  We find that disc stability is strongly modified by the inclusion of turbulent mixing, which pushes the stability threshold to much higher accretion rates.  Thus, turbulent mixing is a viable mechanism for stabilizing the disc at high accretion rates; it is much stronger than convection alone, which has trouble pushing the stability threshold above $\sim 5$ per cent Eddington (compare centre and rightmost tracks in Fig. \ref{fig:scurveSpin}).  The action of turbulence is essentially a stronger version of convection -- the added motion of the turbulent eddies preferentially increases gradients within the disc in the same way that convection does.  The net effect is to further reduce the interior temperature of the gas, thereby requiring the disc to hit much higher accretion rates before it can transition to being radiation dominated (compare the $\dot{M}$ required to reach the same $P_{\rm rad}/P_{\rm gas}$ ratio for the different classes of discs in Fig. \ref{fig:P}).  

\begin{figure}
\begin{center}  
\includegraphics[width=0.5\textwidth]{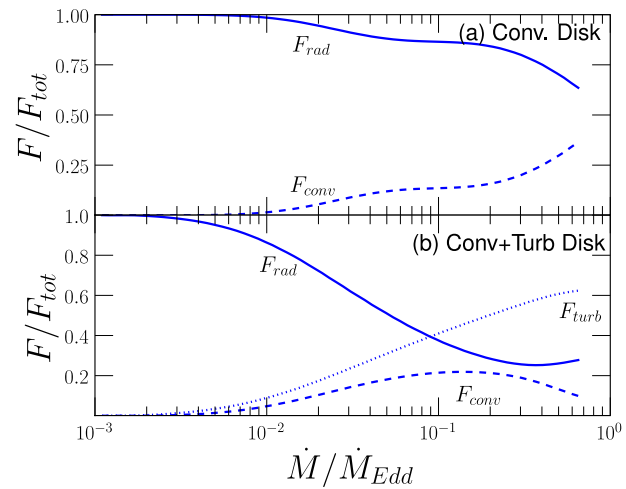}
\caption{Plot of the radiative (solid), convective (dashed), and turbulent (dotted) vertical fluxes as a fraction of the total vertical flux for different accretion rates.  Model parameters: $a_*=0$, $r_*=12$, $\alpha=0.1$, $\Lambda/H=1$.  Panel (a) shows the fluxes in the convective disc, whereas panel (b) shows the results from the convective and turbulent disc.  In both cases, radiation provides the bulk of the vertical energy transport at low accretion rates $\dot{M} < 0.01\dot{M}_{\rm Edd}$.\label{fig:fluxFraction}}  
\end{center}  
\end{figure}

Although we computed solutions all the way up to the Eddington limit, we do not find any cases where convection is overwhelmingly dominant.  In Fig. \ref{fig:fluxFraction}, we see that convection never accounts for more than $\sim 1/2$ of the total vertical energy flux.  This is in line with previous studies on convective discs \citep{shakura78,cannizzo92,milsom94,heinzeller09}.  In turbulent disc solutions, we also find that the turbulent flux $F_{\rm turb}$ becomes dominant at high accretion rates.  Fig. \ref{fig:fluxFraction}(b) shows the partition of turbulent vertical flux in the ``Conv + Turb'' model.  A consequence of this dominant turbulent flux at large $\dot{M}$ is to push down the total radiative flux.  This in turn causes the disc to become cooler overall, which lowers the radiation to gas pressure ratio everywhere (Figs. \ref{fig:P} and \ref{fig:pfracComp}). These cooler discs must reach higher $\dot{M}$ values to hit the same ratio of $P_{\rm rad}/P_{\rm gas}$ and hence require higher $\dot{M}$ to become unstable (since instability is governed by exceeding a critical $P_{\rm rad}/P_{\rm gas}$ ratio).

In addition, there is a second effect in operation.  For the purely radiative standard disc, the transition to instability happens at $P_{\rm rad}/P_{\rm gas}=1.5$.  This is no longer true for convective/turbulent models.  The critical pressure ratio is 2.1 for the convective model and is as large as 7.3 for the convective and turbulent model (compare black dots in Fig. \ref{fig:P}).  This is not too surprising since the critical pressure ratio is achieved when the growth rate for cooling and heating balance (as a function of fluid temperature).  When the convective and turbulent cooling channels are introduced, one must increase the overall cooling rate of the system to keep the same radiative flux as before.  The heating rate must therefore increase to balance this increase in cooling.  The higher heating rate implies higher temperatures, or higher radiation to gas pressure ratios.  Thus, including non-radiative channels causes the critical $P_{\rm rad}/P_{\rm gas}$ to increase.  

The net result of these two effects is that models with both convection and turbulence have critical $\dot{M}$ values well above 10 per cent Eddington.  The cooler interior caused by the suppression of radiative flux, coupled with the increase in the critical $P_{\rm rad}/P_{\rm gas}$ ratio, yields a strong stabilizing effect on the disc.  In the models shown in Fig. \ref{fig:scurveSpin}, the critical $\dot{M}$ for turbulent models is about a factor of 10 higher than that of the purely radiative model.  As we show in \S\ref{sec:sigmaScale} \& \S\ref{sec:zeta}, even larger changes are possible with modest changes to disc parameters.

\subsection{Radial structure of solutions}

\begin{figure}
\begin{center}  
\includegraphics[width=0.5\textwidth]{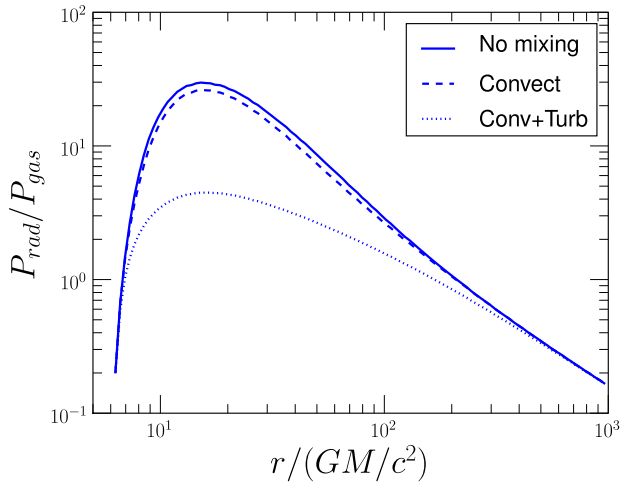}
\caption{Radiation to gas pressure ratio as a function of disc radius for the three classes of disc models.  Model parameters were set to $a_*=0$, $\dot{M}/\dot{M}_{\rm Edd}=0.1$, $\alpha=0.1$, $\Lambda/H=1$, and $\zeta=\alpha$.\label{fig:pfracComp}}
\end{center}  
\end{figure}

\begin{figure}
\begin{center}  
\includegraphics[width=0.5\textwidth]{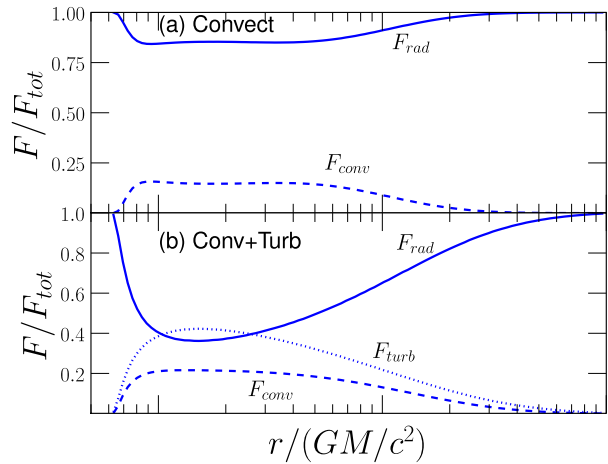}
\caption{Radial dependence of the fractional vertical fluxes in the radiative (solid), convective (dashed), and turbulent (dotted) channels for 10 per cent Eddington solutions.  The upper panel (a) corresponds to the purely convective case, and the lower panel (b) corresponds to the turbulent plus convective case.  Note that the convective flux saturates at about $\sim1/5$ of the total flux.  Model parameters are the same as in Fig. \ref{fig:pfracComp}.\label{fig:fluxComp}}
\end{center}  
\end{figure}

\begin{figure}
\begin{center}  
\includegraphics[width=0.5\textwidth]{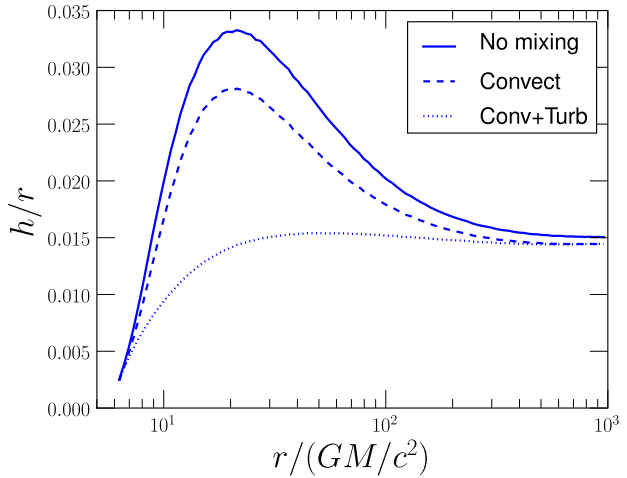}
\caption{Radial profiles of the disc scale height for the three models.  The net result of convective/turbulent mixing is to cool the disc, yielding thinner discs than the purely radiative unmixed disc.  Model parameters are the same as in Fig. \ref{fig:pfracComp}.\label{fig:hComp}}
\end{center}  
\end{figure}

\begin{figure}
\begin{center}  
\includegraphics[width=0.5\textwidth]{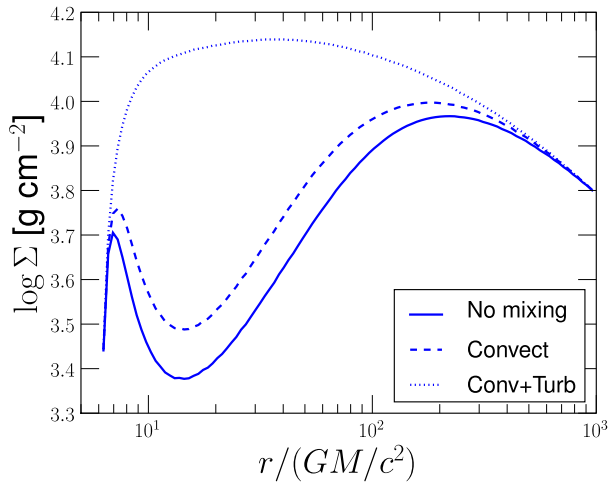}
\caption{Radial dependence of the vertical column mass for the three disc models considered.  Including convective and turbulent mixing causes the vertical column mass to increase everywhere.  Model parameters are the same as in Fig. \ref{fig:pfracComp}.\label{fig:sigmaComp}}
\end{center}  
\end{figure}

Thus far, we have focused on the impact that turbulent energy transport has on the S-curve at a fixed radius.  We now consider the variation in the three disc models with radius.  In all plots below (Figs. \ref{fig:pfracComp} - \ref{fig:sigmaComp}), we show the results for a non-spinning black hole accreting at 10 per cent Eddington.  Other disc parameters are set to their fiducial values (as listed in \S\ref{sec:results}), though the results remain qualitatively the same regardless of the choice of disc parameters.

At large radii, the disc is gas pressure dominated, whereas radiation pressure dominates at smaller radii.  Near $r_*\sim15$, where the bulk of the disc luminosity is released, one finds the highest ratio of $P_{\rm rad}/P_{\rm tot}$ (shown in Fig. \ref{fig:pfracComp}).  At yet smaller radii ($r_*\lesssim8$), the disc transitions back to being gas pressure dominated.  This is due to the luminosity profile (and hence disc temperature) dropping off to zero at $r_*=6$ (the location of the ISCO), which arises from the assumption of zero net torque at the ISCO in the classical theory of accretion discs\citep{pagethorne74}.  Recent analysis of GRMHD simulations of the disc show that although the luminosity plummets sharply at the ISCO, the tenuous plunging gas in the inner region can remain hot and stay radiation dominated \citep{zhu12}.  This effect is beyond the scope of the present work, so we focus on radii $r_* >8$.

From the radial structure of the disc, we find that as mixing becomes more important (as a result of either convection or turbulent mixing), the disc interior becomes cooler.  In models with turbulent/convective modes of vertical energy transport, the hot inner radiation dominated region cools off and becomes more gas-pressure dominated (see the range $10<r_*<100$ in Fig. \ref{fig:pfracComp}, where $P_{\rm rad}/P_{\rm gas}$ shifts towards the gas-limit when mixing is introduced).  This effect occurs due to the additional non-radiative channels for vertical energy transport; less overall energy now travels along the radiative channel(notice the lowering of $F_{\rm rad}$ in panel (b) of Fig. \ref{fig:fluxComp} compared to panel (a) due to the addition of the turbulent flux channel).  This added cooling from mixing also causes the disc to become thinner (compare disc thickness profiles in Fig. \ref{fig:hComp}), which yields higher mass column densities due to the scaling $\Sigma\propto H^{-2}$ at constant flux and radius.  Fig. \ref{fig:sigmaComp} shows the disc column density profiles.  
%


\section{Discussion}\label{sec:discussion}
In developing our simplified disc model, we have made a number of assumptions.  Below we justify our choices, and discuss their impact on the results.

\subsection{Use of logarithmic temperature gradient}  
\begin{figure}
\begin{center}  
\includegraphics[width=0.5\textwidth]{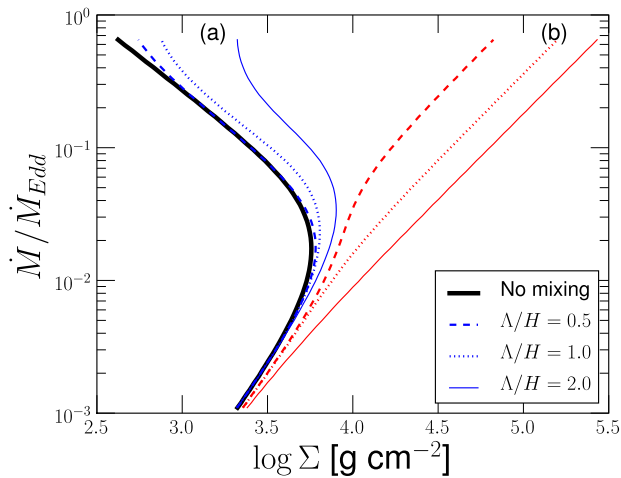}
\caption{S-curves for two different $\nabla$ prescriptions: (a) Blue = $\nabla$ as defined by Eq. \eqref{eq:grad}, (b) Red = alternate definition for $\grad$ according to Eq. \eqref{eq:altgrad}.  For reference we also plot the standard radiative disc solution in thick black.  The red (b) solutions severely overestimate the strength of convection since $\grad$ reaches unphysically large values.  We believe that these solutions correspond to the set of superefficient convective solutions found in some previous studies (e.g. \citealt{goldman95} -- compare with their figure 1). \label{fig:gradMax}}
\end{center}  
\end{figure}

In setting the temperature gradient for our one-zone model, we opted to take the quotient of logarithmic differentials (cf. Eq. \ref{eq:grad}).  Another possibility is to set:
\begin{equation} \label{eq:altgrad}
\grad = \frac{d\ln{T}}{d\ln{P}} \approx \left(\frac{P_{\rm mid}}{T_{\rm mid}}\right) \left(\frac{T_{\rm mid} - T_{\rm scale}}{P_{\rm mid} - P_{\rm scale}}\right),
\end{equation}
which is similar to the prescription used by \citet{goldman95}:
\begin{equation} 
\frac{d\ln T}{d\ln z} \sim \frac{H}{T}\frac{T-T_{s}}{H} = 1-\frac{T_s}{T}.\label{eq:altT}
\end{equation}
 However, taking the logarithm outside of the differential biases $\nabla$ towards large values\footnote{Mathematically, this upwards bias occurs whenever $0<\nabla<1$ for any two variables with scaling ($x\propto y^\nabla$).}.  In particular, when the second probe point is chosen at a scale height, the prescription corresponding to Eqs. \eqref{eq:altgrad} and \eqref{eq:altT} produces unphysically large values of $\grad$.  As a result, convection becomes overwhelmingly strong, with $F_{\rm conv}/F_{\rm tot} \rightarrow 1$ at large $\dot{M}$.  This artefact in $\nabla$ distorts the disc solutions so severely that the disc becomes thermally stable everywhere (compare left and right solution tracks in Fig. \ref{fig:gradMax} corresponding to two prescriptions for $\nabla$).  We believe this choice in evaluating $\nabla$ is why some previous one-zone estimates (e.g. \citealt{bisnovatyi77,goldman95}) found the efficiency of convection to be exceedingly large, in contrast to later more detailed convective disc models that included the full vertical structure and find that the efficiency of convection saturates at $F_{\rm conv}/F_{\rm tot} \sim 1/2$ \citep{cannizzo92,milsom94,heinzeller09}.  Our one-zone model calculates $\grad$ by means of Eq. \ref{eq:grad} and produces results consistent with the latter studies.

\begin{figure}
\begin{center}  
\includegraphics[width=0.5\textwidth]{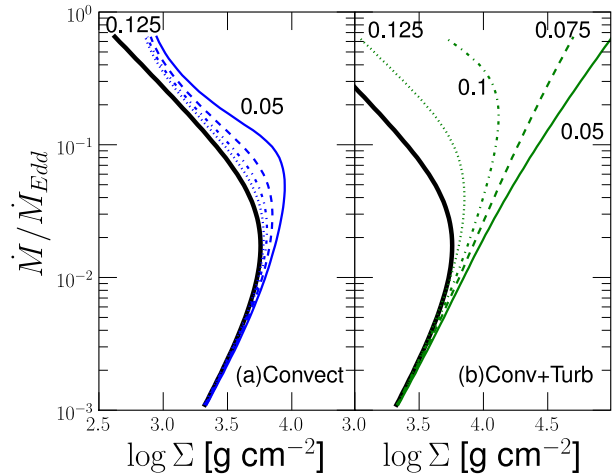}
\caption{S-curves for various choices of $\Sigma_{\rm scale} / \Sigma = (0.125, 0.1, 0.075,0.05)$ shown by (dotted, dash-dotted, dashed, solid) lines respectively.  The thick black line denotes the fiducial S-curve as determined from the purely radiative standard disc. Lines in panel (a) denote the family of convective disc models, whereas those in (b) denote the family of turbulently mixed disc models.  For each choice of $\Sigma_{\rm scale}/\Sigma$, the critical $\dot{M}$ for the turbulent model (b) is much higher than for the corresponding convective model (a).  For some choices of $\Sigma_{\rm scale}/\Sigma$, the turbulent model (b) is completely stable at all $\dot{M}$.\label{fig:scurveScale}}  
\end{center}  
\end{figure}  
\subsection{Impact of $\Sigma_{\rm scale}$}\label{sec:sigmaScale}

In our disc model, one arbitrary parameter is the scale height $\Sigma_{\rm scale}$ used as the upper probe point in defining the temperature gradient $\grad$.  In all models presented so far, we set $\Sigma_{\rm scale}=0.1\Sigma$.  Fig. \ref{fig:scurveScale} shows the impact of varying $\Sigma_{\rm scale}$.  We find that the critical $\dot{M}$ is quite sensitive to the choice of $\Sigma_{\rm scale}$.  In general, larger values of $\Sigma_{\rm scale}$ act to weaken the strength of mixing, resulting in smaller critical $\dot{M}$ values.  However, regardless of the value picked for $\Sigma_{\rm scale}$, the critical $\dot{M}$ for the turbulently mixed disc is always much higher than that of the convective disc (compare left and right panels in Fig. \ref{fig:scurveScale} for the same choice of $\Sigma_{\rm scale}$).

We have looked at more detailed treatments of the vertical structure to see how good our estimate of $\Sigma_{\rm scale}$ is.  For the purely radiative standard disc, we have computed a few detailed models of the vertical structure across a wide range of accretion rates (from $0.1-50$ per cent Eddington) using the stellar atmospheres code TLUSTY \citep{hubeny95,davis05}.  We find $\Sigma_{\rm scale} \sim 0.03-0.15 \Sigma$, with a systematic trend such that $\Sigma_{\rm scale}/\Sigma$ is lower at large $\dot{M}$.  This is simply a consequence of the disc becoming more centrally concentrated when it is cool and in the gas-pressure dominated limit.

As another reference point, a polytropic disc with equation of state $P=K\rho^{1+1/n}$, where $K$ is a constant and $n$ is the polytropic index, has a density profile given by
\begin{equation}
\frac{\rho(z)}{\rho(z=0)}=\left(1-\left(\frac{z}{H}\right)^2\right)^n,
\end{equation} 
where z is the height above the midplane, $H$ is the total height of the disc, and $\rho_0$ is the central density.  For $n=3/2$ (corresponding to a strongly convective column), we have $\Sigma_{\rm scale}/\Sigma \sim 0.04$.  For $n=3$ (corresponding to a radiatively cooled column), we find $\Sigma_{\rm scale}/\Sigma \sim 0.06$.

For the case of an isothermal disc, the vertical density profile is given by
\begin{equation}
\frac{\rho(z)}{\rho(z=0)}=\exp\left[{-z^2/2}\right].
\end{equation}
This corresponds to $\Sigma_{\rm scale} = {\rm erfc(1)}\cdot \Sigma/2 \approx 0.08 \Sigma$.  In all cases, the value of $\Sigma_{\rm scale}$ is pretty close (i.e. within a factor of 2) to the fiducial value that we have adopted in our  models ($\Sigma_{\rm scale}\sim 0.1$).  If anything, our choice is conservative since values of $(\Sigma_{\rm scale} < 0.1 \Sigma)$ enhance the stability of these models (Fig. \ref{fig:scurveScale}).

\subsection{Scaling of Critical $\dot{M}$ with Black Hole Mass}

\begin{figure}
\begin{center}  
\includegraphics[width=0.5\textwidth]{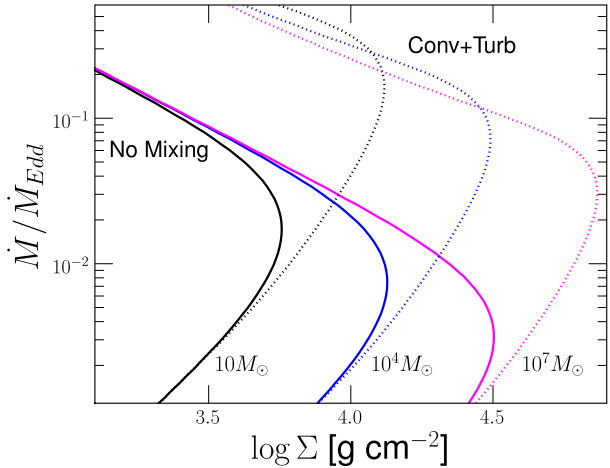}
\caption{S-curve for various choices of black hole masses.  Solid lines denote standard discs without convective/turbulent transport, whereas dotted lines show turbulent + convective disc solutions.  Except for black hole mass, all disc parameters were set to their fiducial values listed in \S \ref{sec:results}. Note that the $M^{-1/8}$ scaling for the critical $\dot{M}$ holds even for the turbulent solutions.\label{fig:scurveM}}  
\end{center}  
\end{figure}

So far, we have only treated the case where $M=10M_{\odot}$.  However, from eq. (2.18) of \citet{shakura73} we expect a weak $\dot{M}_{\rm crit} \propto M^{-1/8}$ scaling for the critical point in the S-curve.  Figure \ref{fig:scurveM} shows the scaling for our disc models.  For supermassive black holes with masses $M \sim 10^{7-8} M_{\odot}$, we expect a roughly tenfold reduction in the critical accretion rate threshold compared to stellar mass black holes, and this is seen in the plot.  Convection and turbulence increase $\dot{M}_{\rm crit}$ by the same factor $\sim 10$.  Even with this, we see that AGN discs are stable only up to luminosities of $\sim 3$ per cent of Eddington.

\subsection{Choosing $\zeta$ -- comparison with simulations}\label{sec:zeta}

\begin{figure}
\begin{center}  
\includegraphics[width=0.5\textwidth]{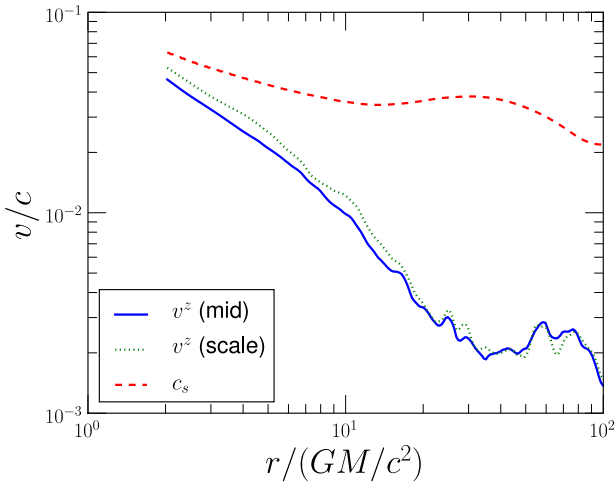}
\caption{GRMHD derived turbulent vertical velocities for a non-spinning black hole from run A0HR07 of \citet{penna10}, in which $\alpha=0.05-0.1$.  The time-averaged root-mean-squared value of the vertical velocity $v^z$ as measured at the disc midplane (blue solid) and at a scale height (green dotted) are shown and compared to the simulation sound speed (red dashed).  The simulation gives $v^z/c_s\sim 0.1$ far from the plunging region (for $r_* > 10$), corresponding to $\zeta\approx 0.1 \approx \alpha$. \label{fig:GRMHDturb}}
\end{center}  
\end{figure}

The value that we take for $\zeta = v_{\rm turb}/c_s$ in our model is primarily motivated by the speed of turbulent motions observed in numerical simulations.  From the global GRMHD simulations of \citet{penna10}, we find that $v_{\rm turb}/ c_s \sim 0.1-0.2$ for $r_*>10$ (compare dashed and solid lines in Fig. \ref{fig:GRMHDturb}).  The effective viscosity coefficient in these simulations is measured to be $\alpha\sim0.05-0.1$, consistent with our choice $\zeta=\alpha$.  However shearing box simulations measure a much larger spread in the vertical advective speed (see figure 23 in \citealt{blaes11}), which can range from $v_{\rm turb} \sim 0.001-0.1 c_s$ depending sensitively on where the advective speed is measured (slowest at midplane, faster at surface).  Note that for shearing box simulations, the effective viscosity $\alpha$ is also much lower than the corresponding values in global simulations; typical values are $\alpha\sim0.01$ \citep{king07}.  Table \ref{tab:zeta} shows a list of inferred $\zeta$ values from various simulations in the literature, and we find the scaling is roughly $\zeta \sim 2 \alpha$.  Note that this scaling for $v_{\rm turb} \propto \alpha$ is inconsistent with an isotropically turbulent $\alpha$-model since the stress would now scale as $W \propto \rho v^2 \propto v_{\rm turb}^2 \propto \alpha^2 \neq \alpha$.  The only way to reconcile this inconsistency is to demand anisotropic turbulence so that the scaling of $W$ with $v_{\rm turb}^2$ no longer holds.

\begin{table}
\caption{Simulation Derived $\zeta=v_{\rm turb}/c_s$ Values}
\begin{center}

\begin{threeparttable}

\begin{tabular}{lccr}
\hline
 Reference & $\alpha$ & $\zeta$ & Method  \\
\hline
\citet{jiang13}  & 0.01-0.02 & 0.036    & Butterfly\tnote{1} \\
\citet{guan11}   & 0.01-0.02 & 0.045    & Butterfly\tnote{1}\\
\citet{blaes11}  & 0.01-0.02 & 0.021    & Direct\tnote{2}\\
\citet{penna10}  & 0.05-0.1  & 0.1-0.2 & Direct\tnote{2} \\
\citet{davis10}  & $\sim$0.01  & 0.026    & Butterfly\tnote{1}\\
\citet{suzuki09} & $\sim$0.01  & 0.025    & Butterfly\tnote{1} \\
\hline
\end{tabular}

\begin{tablenotes}
\item [1] Vertical advection speed is measured from the slope of the simulation butterfly diagram (which shows the characteristic vertical motion of the turbulent dynamo).  This slope yields $v_{\rm turb}$ in terms of $H\Omega = c_s$.  In cases with non-constant slope, we use the slope measured at $z/H=1$.
\item
\item [2] We compare direct measurements of $v_{\rm turb} = v^z$ and $c_s$ at $z/H=1$ to get $\zeta$.

\end{tablenotes}

\label{tab:zeta}
\end{threeparttable}

\end{center}
\end{table}

\begin{figure}
\begin{center}  
\includegraphics[width=0.5\textwidth]{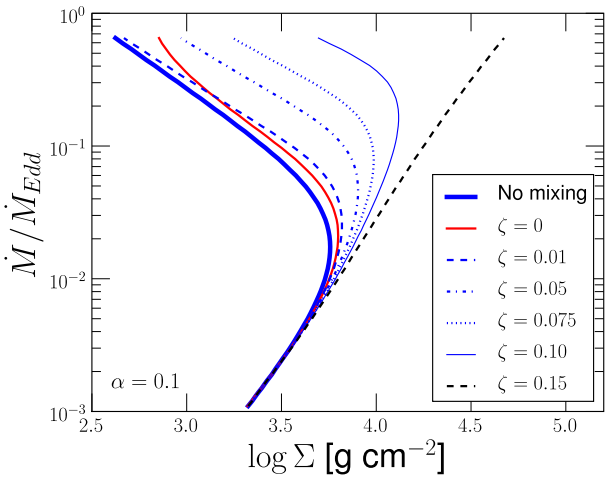}
\caption{S-curve for various choices of the turbulent mixing parameter $\zeta$.  Model parameters: $\alpha=0.1$, $a_*=0$, $r_*=12, \Lambda/H = 1$.  The red solid track ($\zeta=0$) represents a convective disc solution (i.e. no turbulent flux).  For small values of $\zeta\sim 0.01$, turbulence has negligible impact on the disc solutions.  However, for $\zeta \gtrsim 0.05$, the critical $\dot{M}$ is increased substantially, and for $\zeta > 0.1$ the disc is stable for all values of $\dot{M}$.\label{fig:scurveTurb}}  
\end{center}  
\end{figure}

\begin{figure}
\begin{center}  
\includegraphics[width=0.5\textwidth]{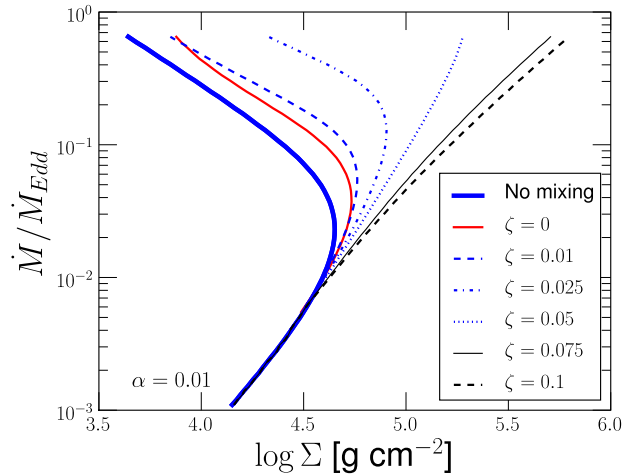}
\caption{S-curve for various choices of $\zeta$ in an $\alpha=0.01$ disc (other parameters identical to those in Fig. \ref{fig:scurveTurb}).  Qualitatively, the behaviour of the solution tracks is similar to the $\alpha=0.1$ case shown in Fig. \ref{fig:scurveTurb}. \label{fig:scurveTurb2}}  
\end{center}  
\end{figure}

To test the sensitivity of our model on the choice of turbulent speed, we now explore a wide range of values for $\zeta$.  The precise location of the critical $\dot{M}$ for stability depends sensitively on the value of $\zeta$ (see the spread of solutions in Figs. \ref{fig:scurveTurb} and \ref{fig:scurveTurb2}).  For sufficiently large $\zeta$ (not much larger than our canonical value of 0.1), the turbulent disc becomes stable for all accretion rates.  Due to sensitivity to model parameters, this prediction should be treated with caution and must be interpreted more as a proof of concept.  Our main result is that turbulent mixing can provide a much stronger stabilizing force than that provided by convection alone.  This can be understood on the basis that MRI-induced turbulence is likely much stronger than convective turbulence, and hence should have a stronger impact on the disc physics.

\subsubsection{Comparison with shearing box simulations}
The state-of-the-art in accretion disc modeling are the detailed vertically stratified Radiation-MHD shearing box simulations of \citet{hiroseblaes09}, \citet{hirosekrolik09} and \citet{blaes11}.  By comparing and analyzing a sequence of these simulations with different initial conditions, \citet{hiroseblaes09} were able to piece together a simulation derived S-curve.  Their resultant S-curve was very similar to the prediction from standard purely radiative disc theory, with one major difference; their radiation pressure dominated branch of solutions was apparently found to be thermally stable.  They attribute this stability to non-synchronous evolution of stress and pressure in the simulations\citep{hirosekrolik09}.  Stress fluctuations were found to precede pressure fluctuations, thereby breaking the usual argument for thermal instability (since the dissipative heating rate $Q^+$ due to stress is no longer set by the pressure, which removes the steep temperature dependence of $Q^+$ in the radiation limit).  These results are in contrast to our turbulent disc model, where we find the S-curve to be significantly modified by the action of turbulent mixing.  

We reconcile this apparent contradiction with the fact that in the shearing box simulations, turbulent mixing is weak.  Looking at the \citealt{blaes11} entry in Table \ref{tab:zeta}, we see their simulation yields a low $\zeta\approx 0.02$.  If we adopt a small enough value for $\zeta$, our model too gives an S-curve that is almost unmodified from the standard-disc S-curve (compare how similar the ``No mixing'', $\zeta=0$, and $\zeta=0.01$ tracks are in Figs. \ref{fig:scurveTurb} and \ref{fig:scurveTurb2}).  One qualitative difference remains between our model and shearing box discs; our model has a thermally unstable upper radiative branch whereas in the simulations of \citet{hiroseblaes09}, this branch appears to be thermally stable.  However, recent work by \citet{jiang13} using a code based on Athena \citep{stone08,jiang12}, indicates that the radiation branch is in fact unstable.  The reason for the difference between the two studies is not understood.

The weak turbulent mixing seen in shearing box simulations may simply be a consequence of their small value of $\alpha\sim0.01$.  Since discs with larger values of $\alpha$ ought to produce larger turbulent velocities, we conjecture that once $\alpha$ becomes sufficiently large (say close to the values observed in real discs, $\alpha \rightarrow 0.1$)  turbulent mixing becomes strong, enabling large changes to occur in the S-curve.  Thus, although shearing box simulations (small $\alpha$) predict no modification to the standard S-curve, it is possible that nature (large $\alpha$) admits S-curves that are significantly modified due to the presence of stronger turbulent motions.

One final distinction between our model and the shearing-box simulations is that \citet{blaes11} find their solutions to be convectively stable everywhere.   They do see vertical advection of energy, but it is entirely due to magnetic buoyancy.  In contrast, our model for turbulent mixing requires a convectively unstable entropy gradient to achieve a net outwards flux (i.e. for positive $F_{\rm turb}$, we require positive $\Delta T$ in Eq. \eqref{eq:fluxturb} which can only occur when $\grade > \grad$).  There is no way to reconcile this difference as our model assumes active convection whenever there is outwards turbulent flux.  It is perhaps too demanding to ask our highly simplified 1-zone disc model to exactly match the results of detailed 3D MHD simulations.

\subsection{$\zeta$ from observations}\label{sec:zetaobs}

Spectral state transitions in stellar mass black hole systems may offer a clue in what nature chooses for $\zeta$.  These transitions are brought about by changes in mass accretion rate of the system\citep{mcclintockremillard06}.  Of particular interest is the transition from the thermally-dominant high/soft state to the steep-power-law very-high state.  The high/soft state, believed to be thermally stable, exhibits little variability and spans the luminosity range 2 - 50 per cent of Eddington \citep{gierlinski04, gierlinski06}.  The very-high state, occurring at yet higher accretion rates, is usually accompanied by significant variability including the emergence of high frequency quasi periodic oscillations.  

The interpretation from our disc models is that this state transition occurs as the disk migrates from the thermally stable lower branch to the thermally unstable upper branch of solutions.  This critical accretion rate is intimately linked to the value of $\zeta$ (c.f. Figures \ref{fig:scurveTurb}, \ref{fig:scurveTurb2}).  Based on observations of the high/soft to very-high state transition, we infer the critical accretion rate to be $\dot{M}_{\rm crit}/\dot{M}_{\rm Edd} \sim 0.5$ \citep{mcclintockremillard06}.  According to our models, this value for $\dot{M}_{\rm crit}$ suggests $\zeta \sim 0.1$ or larger in our fiducial disc model.  We are unable to make a very precise prediction regarding $\zeta$ since the value of $\dot{M}_{\rm crit}$ in our model is also affected by the other disc parameters.

\subsection{Radiative Outer Zone}

The vertical transport of energy at the disc surface must be dominated by radiative flux (e.g. \citealt{blaes11} found that the radiative flux dominates over advective flux above a few disc scale heights in shearing-box simulations).  In our one-zone model, we assume that this radiative outer zone is thin and can be neglected.  However, this assumption produces a systematic bias in the disc solutions; the presence of a secondary purely radiative zone acts to increase the interior temperature of the disc.  This can be understood on the basis of the radiative diffusion equation.  A purely radiative zone has a larger $F_{\rm rad}$ than an equivalent model with convective and turbulent mixing.  According to Eq. \eqref{eq:radiativediffusion}, a consequence of this larger $F_{\rm rad}$ is a larger interior temperature.

\begin{figure}
\begin{center}  
\includegraphics[width=0.5\textwidth]{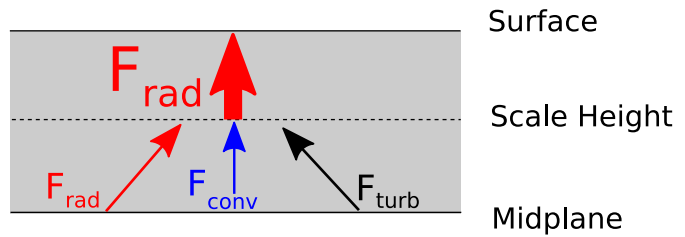}
\caption{A schematic of the vertical structure in our \emph{two-zone} model (compare to the one-zone model in Fig. \ref{fig:schematic}).  We assume that the region between the surface and the density scale height is purely radiative.  The interior is identical to that of our previous one-zone models.\label{fig:schematic2}}  
\end{center}  
\end{figure}

\begin{figure}
\begin{center}  
\includegraphics[width=0.5\textwidth]{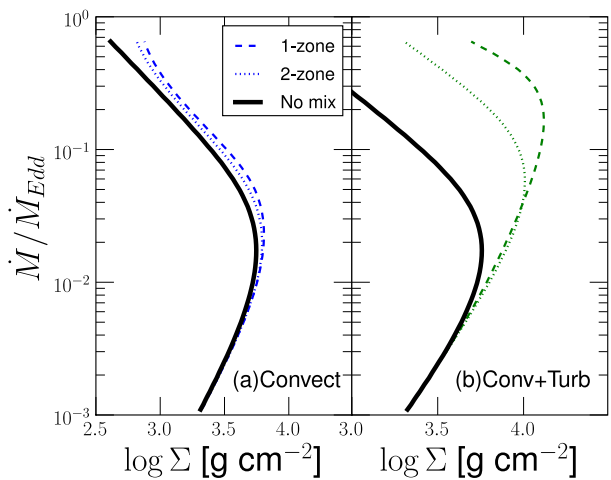}
\caption{S-curves of the two-zone model (dotted) compared to the corresponding one-zone model(dashed) and purely radiative standard disc (solid).  Note that the critical $\dot{M}$ for turbulent discs (right panel) are a factor $\sim 4-5$ larger than in purely convective discs (left panel).  Model parameters: $a_*=0$, $r_*=12$, $\alpha=0.1$, $\Lambda/H=1$, and $\zeta=\alpha$\label{fig:scurveF}}  
\end{center}  
\end{figure}

The exact location of the boundary separating the convective and turbulent interior from the purely radiative surface layer depends crucially on the details of the disc's vertical structure.  Since we only have a one-zone model to work with, we make the following crude and extreme assumption:  we assume that the density scale height is the demarcation point between a convective and turbulent interior and a purely radiative exterior (Fig. \ref{fig:schematic2}).
This jump to two-zones primarily changes the value of $T_{\rm scale}$.   Since in the radiative zone $F_{\rm rad} = F_{\rm tot} = \sigma T_{\rm eff}^4$, the previous radiative diffusion equation (Eq. \ref{eq:radiativediffusion}) now becomes
\begin{equation}
T(\tau) = T_{\rm eff} \cdot \left[\frac{3}{4}\left(\frac{2}{3} + \tau\right)\right]^{1/4}. \label{eq:radiativediffusionNew}
\end{equation}
For simplicity, we do not modify the other disc equations so the system can be solved in the same way  our one-zone models.  Fig. \ref{fig:scurveF} compares the one-zone and two-zone disc solutions.  As expected, the inclusion of a radiative outer zone pushes the convective and turbulent solutions towards that of a purely radiative standard disc (i.e. the two-zone solution lies in between the purely radiative disc and the homogeneously mixed one-zone disc).  The radiative outer zone ultimately pushes down the critical $\dot{M}$ of the two-zone model towards lower values.  However, our main result is unaffected; even in two-zone discs, turbulently mixed discs are still significantly more stable than their purely convective counterparts (compare critical $\dot{M}$ of left and right panels in \ref{fig:scurveF}).

\subsection{Complete stabilization from turbulence}

\begin{figure}
\begin{center}  
\includegraphics[width=0.5\textwidth]{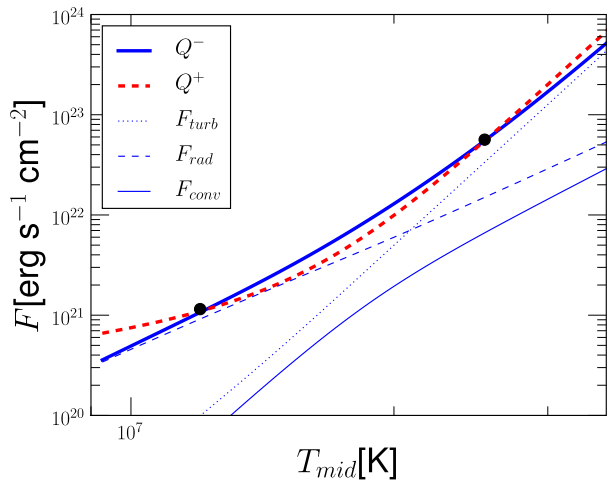}
\caption{Comparison of cooling flux $Q^-$ (blue, thick solid) with heating flux $Q^+$ (red, thick dashed) for turbulent disc solutions with $\zeta=0.1$, $\alpha=0.1$, $r_*=12$, $a_*=0$, $\Lambda/H=1$, and $\log \Sigma=3.8$.  We further break down the cooling flux into its various components: $F_{\rm turb}$ (thin dotted), $F_{\rm rad}$ (thin dashed), and $F_{\rm conv}$ (thin solid).  The two black dots denote the two stationary disc solutions where cooling matches heating (i.e. the two points on the S-curve corresponding to the selected column density).  The lower dot represents a stable solution and the upper dot an unstable solution.\label{fig:QcQh-unstab}}
\end{center}  
\end{figure}

\begin{figure}
\begin{center}  
\includegraphics[width=0.5\textwidth]{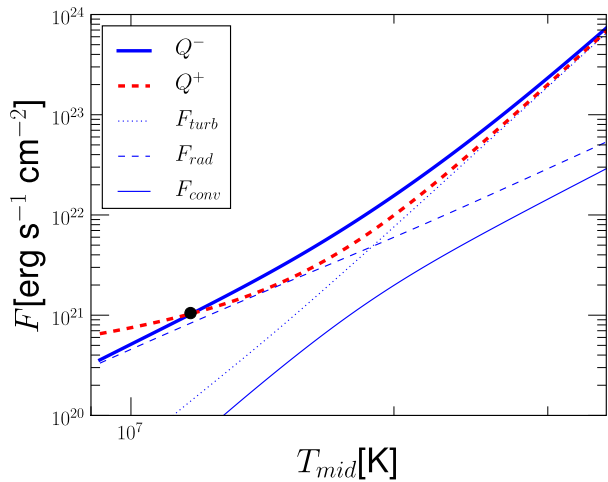}
\caption{Same as Fig. \ref{fig:QcQh-unstab}, but for a model with stronger turbulent mixing ($\zeta=0.15$).  Here, the disc solution is completely thermally stable, so there is only one stationary disc solution (black dot, which is thermally stable).  The stability is due to the strong turbulent cooling flux which ensures that $Q^- > Q^+$ at large $T_{\rm mid}$.\label{fig:QcQh-stab}}
\end{center}  
\end{figure}

For certain choices of model parameters (i.e. small $\Sigma_{\rm scale}$ and/or large $\zeta$), we find that our turbulent and convective disc models admit solutions that are thermally stable at all accretion rates.  Solutions of this form exhibit a linear track in the $\dot{M} - \Sigma$ plane with positive slope at all $\dot{M}$ (see the rightmost track in Figs. \ref{fig:scurveScale}, \ref{fig:scurveTurb}, and \ref{fig:scurveTurb2}).  To understand what qualitative differences exist between solutions that have a critical transition point and those that are completely stable everywhere, we examine the heating and cooling curves for these two scenarios (compare Fig. \ref{fig:QcQh-unstab} which exhibits both a stable and unstable disc solution, and Fig. \ref{fig:QcQh-stab} where only a single stable solution exists).  In these heating/cooling curves, we hold fixed the total vertical column density and solve our usual set of disc equations for various choices of disc midplane temperature.
We find for our turbulence model that the cooling scales as $F_{\rm turb} \sim T^8$ -- comparable to the scaling for heating in the radiation dominated regime $Q^+ \sim T^8$.  If the action of turbulence becomes sufficiently strong (represented in our model by taking a large value for $\zeta$), then at high temperatures cooling always overtakes heating (see Fig. \ref{fig:QcQh-stab}).  In these strongly turbulent cases, any positive temperature perturbations above the equilibrium solution eventually cool back down to equilibrium.  This eliminates the upper unstable branch of solutions, leading to complete stability.


\section{Summary}\label{sec:conclusion}

In this work, we have developed a simple one-zone model for black hole accretion discs with both convective and turbulent vertical energy transport.  We find that the action of mixing from convection and turbulence provides a stabilizing effect on the disc, pushing the threshold for thermal instability up towards higher accretion rates (compare the critical $\dot{M}$ for different classes of discs in Fig. \ref{fig:scurveSpin}).  For stellar mass black holes, convection by itself provides only a modest boost to the thermal instability threshold, only pushing the critical $\dot{M}$ to 5 per cent Eddington in the most favourable cases.  On the other hand, models that include additional mixing through MRI-induced turbulence are much more stable.  In some cases, we find that turbulent mixing pushes the threshold for instability far above 10 per cent Eddington -- even inducing complete stability in the most extreme cases.  A similarly strong effect is seen for supermassive black holes.  However, since the critical $\dot{M}$ for stability is lower by a factor of $\sim10$, even with the effect of convection and turbulence, stable solutions are found only up to a few per cent of Eddington.\\

Previous studies have shown that convective mixing in discs tends to provide a stabilizing effect \citep{cannizzo92,milsom94}, which raises the critical accretion rate marking the onset of thermal instability.  Since MRI-induced turbulence is much more vigorous than convective turbulence, it is not surprising that turbulently mixed discs experience a much stronger version of this stabilizing effect.  Thus, we believe that thermal stabilization from turbulent mixing is a promising mechanism for explaining the apparent lack of instability in luminous (up to 50\% of Eddington) black hole X-ray binaries \citep{gierlinski04}.  However, due to the mass scaling for the critical accretion rate, we do not expect supermassive black holes to be stable at such high accretion rates - in our models they become unstable at around a few per cent of Eddington.\\

Finally, we stress that the precise value for the critical $\dot{M}$ found in our models should not be taken too seriously.  We are working in the framework of a one-zone model (a rather severe approximation), and thus the quantitative details of our model are quite sensitive to the choice of model parameters.  Our main result is a qualitative one -- MRI-induced turbulent discs are much more stable than equivalent convective discs.\\

\section{Acknowledgments}

The authors would like to thank Robert Penna, Aleksander S\c{a}dowski, and Dmitrios Psaltis for insightful discussions about disc convection.  We thank the anonymous referee for providing illuminating comments and helping us craft a much clearer paper.  YZ also thanks Tanmoy Laskar for helpful suggestions for improving the presentation of the manuscript.  YZ was supported by the Smithsonian Institution Endowment Funds.  This work was supported in part by NASA grant NNX11AE16G.\\


\bibliography{ms}

\appendix

\section{Solving for 8 unknowns}\label{app:solvingEqs}

Due to the highly non-linear nature of Eqs. \eqref{eq:pressurebal}-\eqref{eq:gamma}, we solve the system of equations numerically.  The technique is as follows:

\begin{enumerate}[(i)]
\item Assume an exploratory trial value for $P_{\rm mid}$ (i.e. we guess at its value, and check at the end if it produces a result that is consistent with all 8 equations).
\item From Eqs. \eqref{eq:pressurebal} and \eqref{eq:heating}, we obtain a relation for $\Sigma$ in terms of $P_{\rm mid}$: 
\begin{equation}
\Sigma = \left(\frac{3\Omega_k R_F \alpha}{\sigma T_{\rm eff}^4 Q}\right)\cdot P_{\rm mid}^2.
\end{equation}
\item Again from Eqs. \eqref{eq:pressurebal} and \eqref{eq:heating}, we also obtain $H$ by plugging in our value of $\Sigma$:
\begin{equation}
H=\sqrt{\frac{4\sigma T_{\rm eff}^4}{3 \Sigma \Omega_k Q R_F \alpha}}\label{eq:Hscaling}.
\end{equation}
\item From Eq. \eqref{eq:midplaneEOS} and given the values of $\Sigma$, $P_{\rm mid}$, and $H$, we can solve for $T_{\rm mid}$ (numerically, by Newton's method).
\item Plugging $\Sigma$ into Eq. \eqref{eq:radiativediffusion}, we immediately get $T_{\rm scale}$.
\item Using $T_{\rm scale}$, $\Sigma$, $H$ in Eq. \eqref{eq:scaleEOS} yields $P_{\rm scale}$.
\item Given all these quantities, using the two Eqs. \eqref{eq:convectivediffusion} and \eqref{eq:gamma} allows us to solve for ($\grad$ - $\grade$).  In particular, we rearrange Eq. \eqref{eq:fluxconservation} to first yield:
\begin{equation}\label{eq:gradeqs}
A(\grad - \grade)^{3/2} + Z(\grad - \grade) = (\gradr - \grad),
\end{equation}
using the following definitions for the gradients $\grad$, $\gradr$:
\begin{equation}\label{eq:gameq1}
F_{\rm tot} \equiv \left(\frac{4ac}{3}\frac{QHT_{\rm mid}^4}{\kappa P_{\rm mid}}\right) \cdot \gradr,
\end{equation}
\begin{equation}\label{eq:gameq}
F_{\rm rad} \equiv \left(\frac{4ac}{3}\frac{QHT_{\rm mid}^4}{\kappa P_{\rm mid}}\right) \cdot \grad,
\end{equation}
and the prefactors:
\begin{equation}
A = \frac{3 \kappa \Sigma P_{\rm mid} c_p}{16\sqrt{2} ac \sqrt{Q} H T_{\rm mid}^3}\left(\frac{\Lambda}{H}\right)^2,
\end{equation}
\begin{equation}
Z = \frac{3 \Sigma c_p \bar{v}_{\rm turb} \kappa P_{\rm mid}}{8ac Q H^2 T_{\rm mid}^3}\left(\frac{\Lambda}{H}\right).
\end{equation}
We eliminate $\grad$ from the RHS of \eqref{eq:gradeqs} by adding $(\grad-\grade)+(\grade-\grada)$ to both sides:
\begin{equation}\label{eq:gradeqs2}
A(\grad - \grade)^{3/2} + (Z+1)(\grad - \grade) + (\grade-\grada) = (\gradr - \grada).
\end{equation}
Now, Eq. \eqref{eq:gammafull} allows us to convert the ($\grade - \grada$) term into a pure function of ($\grad - \grade$), giving the quadratic:
\begin{equation}\label{eq:gameq2}
\frac{(\grad - \grade)^2}{(\grade-\grada)^2} - \left(\frac{2\eta \bar{v}_{\rm turb}}{\grade - \grada} + \frac{\eta^2 \Lambda^2 Q}{8}\right)\cdot(\grad - \grade) + \eta^2 \bar{v}_{\rm turb}^2 = 0,
\end{equation}
where:
\begin{equation}
\eta = \left(\frac{\Sigma c_p}{16 \sigma T_{\rm mid}^3 H}\right) \cdot \left(\frac{1+\tau_m^2/2}{\tau_m}\right).
\end{equation}
We take the positive root for ($\grade-\grada$) in Eq.\eqref{eq:gameq2}, and plug it into Eq.\eqref{eq:gradeqs2}.  This produces a polynomial expression for ($\grad - \grade$), which can be solved numerically.

In the limiting case where $\bar{v}_{\rm turb} \rightarrow 0$ (case of pure convection without additional turbulent mixing), we find that Eq. \eqref{eq:gameq2} simplifies to:
\begin{equation}
(\grade-\grada) = B (\grad - \grade)^{1/2} \label{eq:gamma-B}
\end{equation}
where 
\begin{equation}
B=\left(\frac{32\sqrt{2}\sigma T_{\rm mid}^3}{\Sigma c_p\sqrt{Q}}\right)\left(\frac{\Lambda}{H}\right)^{-1}\left(\frac{\tau_m}{1+\tau_m^2/2}\right).\label{eq:B}
\end{equation}
Furthermore, $\bar{v}_{\rm turb} \rightarrow 0$ also implies $Z \rightarrow 0$ and Eq.\eqref{eq:gradeqs2} becomes the following cubic, which can be easily solved for $(\grad - \grade)$:
\begin{equation}\label{eq:gradeqs3}
A(\grad - \grade)^{3/2} + (\grad - \grade) + B(\grad-\grade)^{1/2} = (\gradr - \grada).
\end{equation}
\item Plugging ($\grad - \grade$) back into Eq. \eqref{eq:gamma} allows us to solve for both $\grad$ and $\grade$ individually.
\end{enumerate}

Now, armed with values for all 8 unknowns, we check to see if we made the correct guess for $P_{\rm mid}$ by checking for consistency in $\grad$.  A correct guess for $P_{\rm mid}$ will cause $\grad$ computed from the two probe points in Eq. \eqref{eq:grad} to be consistent with $\grad$ computed from step (viii).  Empirically, we find that this relation for $\grad$ is monotonic with $P_{\rm mid}$, and hence we are able to solve for the correct value of $P_{\rm mid}$ via bisection.


\end{document}